\documentclass[prl,twocolumn,floatfix,superscriptaddress]{revtex4-2}

\usepackage{amsmath, amssymb}
\usepackage{enumitem}
\usepackage{mathtools}
\usepackage{lipsum}
\usepackage{times}
\usepackage{graphicx}
\usepackage{wasysym}
\usepackage{bm}
\usepackage{hyperref}
\usepackage{xspace}
\usepackage{siunitx}
\usepackage{xcolor}
\usepackage{soul}
\usepackage{braket}
\usepackage[normalem]{ulem}

\definecolor{myColor}{rgb}{0.02,0.12,0.8}%0.3}
\definecolor{myciteColor}{rgb}{0.39,0.7,0.89}
\hypersetup{colorlinks=true, linkcolor=myColor, filecolor=myColor, urlcolor=myColor,citecolor=myColor, urlcolor=myColor}
\graphicspath{{./Figures/}}

\DeclareSIUnit{\nK}{\nano\kelvin}
\DeclareSIUnit{\aB}{\emph{a}_0}
\DeclareSIUnit{\G}{G}

\renewcommand{\figurename}[1]{Fig.~}
\newcommand{\kB}{k_{\text{B}}}

\newcommand{\lambdac}{\lambda_\mathrm{c}}

\begin{document}

\title{Fluctuation-dissipation relation for a Bose-Einstein condensate of photons}

\author{Fahri Emre \"Ozt\"urk}
\email{oeztuerk@iap.uni-bonn.de}
\author{Frank Vewinger}
\author{Martin Weitz}
\author{Julian Schmitt}
\email{schmitt@iap.uni-bonn.de}

\affiliation{Institut f\"ur Angewandte Physik, Universit\"at Bonn, Wegelerstr. 8, 53115 Bonn, Germany}

\date{\today}

\begin{abstract}
For equilibrium systems, the magnitude of thermal fluctuations is closely linked to the dissipative response to external perturbations. This fluctuation-dissipation relation has been described for material particles in a wide range of fields. Here we experimentally probe the relation between the number fluctuations and the response function for a Bose-Einstein condensate of photons coupled to a dye reservoir, demonstrating the fluctuation-dissipation relation for a quantum gas of light. The observed agreement of the scale factor with the environment temperature both directly confirms the thermal nature of the optical condensate and demonstrates the validity of the fluctuation-dissipation theorem for a Bose-Einstein condensate.
\end{abstract}


\maketitle
The fluctuation-dissipation theorem, relating the thermal fluctuations of a system at temperature $T$ to its response to an external perturbation by the thermal energy $\kB T$, is a cornerstone of statistical mechanics~\cite{Callen:1951,Kubo:1966}. Experimentally, it has been observed in a wide range of systems, {\it e.g.}, with particles undergoing Brownian motion~\cite{Einstein:1905}, the statistical fluctuations of electrical currents in resistors~\cite{Johnson:1928}, and more recently also in cold atomic gas settings~\cite{Esteve:2006,Mueller:2010,Sanner:2010}, including two-dimensional Bose superfluids in the strongly interacting regime~\cite{Gemelke:2009,Hung:2011}. The relation provides an elegant approach to access microscopic properties of a system (fluctuations) by probing the response on a macroscopic level (dissipation), allowing to determine equilibrium quantities such as the structure factor, which would be difficult to access otherwise~\cite{Blumkin:2013,Christodoulou:2021}.

For Bose-Einstein condensates, despite that this phase is one of the most thoroughly investigated quantum states of matter, the fluctuation-dissipation theorem could so far not be examined. In cold-atom condensates thermal number fluctuations are strongly suppressed~\cite{Kristensen:2019,Christensen:2021}, and in optical condensates the driven-dissipative nature of such systems~\cite{Bloch:2022} has kept the possibility for a successful test of the fluctuation-dissipation relation an open question. Interestingly, the fluctuation-dissipation relation can be extended to non-equilibrium systems in steady state such as lasers~\cite{Agarwal:1972}, however there the scaling with temperature -- a universal quantity -- is replaced by system-specific two-point correlation functions~\cite{Haenggi:1975,Prost:2009}. Along this line, theory work has recently pointed out that probing the validity of the fluctuation-dissipation relation provides a very direct and critical test of thermalization and allows one to characterize the eventual departure from equilibrium in optical quantum gases~\cite{Chiocchetta:2017b}.

A new approach to study fluctuations and the corresponding response function in the condensed phase has emerged in quantum gases as exciton-polaritons and photons, where a coupling to reservoirs is realized~\cite{Askitopoulos:2013,Schmitt:2014,Schmitt:2016,Baboux:2018,Pieczarka:2020,Wang:2021}. 
In the latter experiments using photons, other than for the case of a blackbody gas, a thermodynamic phase transition to a Bose-Einstein condensate can be observed, {\it e.g.}, in two-dimensional dye-filled optical microcavity systems~\cite{Klaers:2010,Marelic:2015,Greveling:2018}. Thermalization here is achieved by absorption-re-emission processes on dye molecules, which provide both an energy and a particle reservoir due to the possible interconversion of cavity photons and dye electronic excitations. This situation can be described by a grand canonical ensemble model, a physical setting for which unusually large fluctuations occur in the condensed phase~\cite{Ziff:1977,Klaers:2012,Sobyanin:2012,Gladilin:2020a}. Experimentally, the corresponding number fluctuations have been observed in the dye microcavity system, with the magnitude of fluctuations being tuneable by adjusting the relative size of the condensate and the effective reservoir~\cite{Schmitt:2014,Schmitt:2016}.

In this letter, we report a measurement of both the spontaneous number fluctuations and the associated reactive response of a photon Bose-Einstein condensate coupled to a reservoir, demonstrating the validity of the fluctuation-dissipation theorem for a Bose-Einstein condensate. By tuning the reservoir size we find that the relation applies from canonical through to grand canonical conditions. Within experimental uncertainties, the observed scaling between fluctuations and the response is consistent with $\kB T$, where $T\approx \SI{300}{\kelvin}$ is the temperature of the reservoir. Such a critical test of the thermalized nature of an optical condensate as well as its coupling to the reservoir goes beyond earlier work that has verified, {\it e.g.}, spectral properties and spatial redistribution of light in trapping potentials~\cite{Deng:2006,Klaers:2010,Schmitt:2015,Sun:2017}, and is also of interest for thermometry in complex lattice or quenched systems~\cite{Cugliandolo:2011,Foini:2011,Foini:2012}.

%%%%%%%%%%%%%%%%%
%%%%% FIG 1 %%%%%
%%%%%%%%%%%%%%%%%
\begin{figure}[t]
    \centering
    \includegraphics[width=1.0\columnwidth]{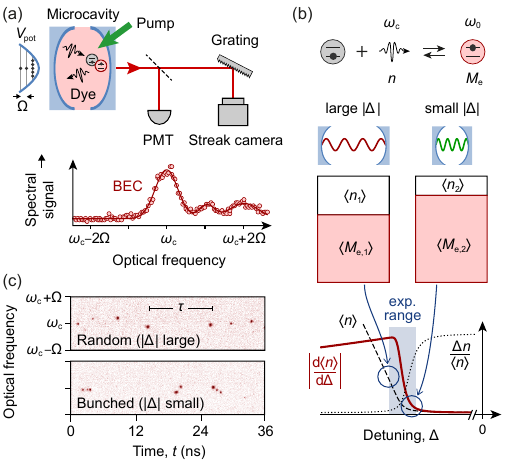}
\caption{(a) Experimental scheme to measure the number fluctuations and the response function of a photon Bose-Einstein condensate coupled to a reservoir inside a dye-microcavity. Part of the cavity emission recorded with a photomultiplier (PMT) yields the mean condensate population $\langle n \rangle$; the other part is dispersed on a grating, and the spectrally filtered condensate evolution is recorded with a streak camera, giving $g^{(2)}(\tau)$ and the dye-cavity detuning $\Delta$. Bottom: time-integrated spectrum showing the condensate mode at $\omega_\mathrm{c}$, well separated from the first excited states in the harmonic potential with trapping frequency $\Omega$. (b) The reservoir is realized by $M_\mathrm{e}$ excited molecules (energy per molecule $\simeq\hbar\omega_0$), coupled to $n$ condensate photons (energy per photon $\hbar\omega_\mathrm{c}$) by interconversion (top). The cavity length controls $\Delta$, adjusting the ratio between photons and excited molecules (middle). Bottom: When detuning from resonance, the predicted photon number and the response (red) sharply increase; at this point the system starts to minimize its free energy by creating photons instead of maximizing the entropy in the molecular reservoir (see text). At large negative detunings $\Delta\rightarrow -\infty$, where $\langle n\rangle\rightarrow X$, the response gradually falls off to 0. (c) Spectrally resolved streak camera traces showing random arrival times of coherent photons for large $|\Delta|$ (top), and bunched photons at small $|\Delta|$ (bottom).}
    \label{fig:1}
\end{figure}

Our photon Bose-Einstein condensates are prepared in a microcavity apparatus shown in Fig.~\ref{fig:1}(a), realized by two curved mirrors filled with a dye molecule solution of refractive index $\tilde{n}=1.44$; see refs.~\cite{Klaers:2010,Schmitt:2018} for details. The cavity length $D_0=q\lambdac/2\tilde n\simeq \SI{1.4}{\micro\meter}$ on the order of the optical wavelength $\lambdac\approx \SI{575}{\nano\meter}$ at mode number $q=7$ introduces a low-energy cutoff at $\hbar \omega_\mathrm{c} \simeq \SI{2.1}{\electronvolt}$, with $\omega_\mathrm{c}=2\pi c/\lambdac$, speed of light $c$ and the reduced Planck's constant $\hbar$. In the microcavity, the photon dispersion relation becomes two-dimensional and matter-like, {\it i.e.}, energy scales quadratically with the wave vector; owing to the mirror curvature, the photons are harmonically trapped with frequency $\Omega/2\pi\simeq \SI{40}{\giga\hertz}$~\cite{Klaers:2010}. Other than, {\it e.g.}, in cold atoms or exciton-polaritons, equilibration of the ensemble does not occur by interparticle collisions, but by contact to a heat bath of photo-excitable dye molecules at room temperature. Thermalization is effective when the contact to the dye molecules dominates over losses, {\it e.g.}, from mirror transmission~\cite{Schmitt:2015}, and the average photon number is controlled by a laser beam pumping the dye. Above a critical photon number, a Bose-Einstein condensate forms in the cavity ground state~\cite{Klaers:2010,Marelic:2015,Greveling:2018}. Absorption and emission processes frequently convert the condensed photons into dye excitations (and vice versa), leading to statistical fluctuations of the condensate number. Thus, the dye acts as a heat bath {\it and also} as a particle reservoir, which enables tuning between a canonical and a grand canonical ensemble description, see Fig.~\ref{fig:1}(b).

The predicted number fluctuations of a photon condensate coupled to a reservoir of $M$ dye molecules and its response function can be calculated from a statistical approach~\cite{Klaers:2012,Sobyanin:2012,Schmitt:2018}. Here the detuning $\Delta=\omega_\mathrm{c}-\omega_0$ between the condensate and the zero-phonon line of the dye $\omega_0\simeq 2\pi c/(\SI{545}{\nano\meter})$ serves as a control parameter; for our experiment, $\Delta<0$. The exchange between photons and dye excitations, see Fig.~\ref{fig:1}(b) (top), preserves the total number of excitations $X=n+M_\mathrm{e}$ with $n$ condensate photons and $M_\mathrm{e}=\sum_{i=0}^M {f_i}$ dye excitations; here, $f_i=\{0,1\}$ refers to the $i$-th dye molecule being in the electronic ground $(0)$ or excited state $(1)$, respectively. This assumption is well justified due to the small overlap of the excited molecules in the ground mode volume with the higher-lying photon modes in the harmonic trap, such that a reservoir-mediated cross-coupling is in general weak, and due to the spatially near-uniform excitation level, such that (on average) excitations do not flow to or away from the ground mode reservoir. In thermodynamic equilibrium, one obtains the photon number probability distribution for $n$ particles in the Bose-Einstein condensate
\begin{equation}
    \mathcal{P}_n=\frac{1}{Z} \frac{M!}{(M-X+n)!(X-n)!} e^{-n\frac{\hbar\Delta}{\kB T}},
\label{eq:1}
\end{equation}
where the partition function $Z$ is determined by normalization $\sum_{n}{\mathcal{P}_n}=1$. Note that eq.~\eqref{eq:1} is derived by assuming thermal contact of the total system of photons and dye molecules to a heat bath (the solvent). Microscopically, the detuning dependence of $\mathcal{P}_n$ is understood from the Kennard-Stepanov relation $B_\mathrm{abs}/B_\mathrm{em} = \exp(\hbar\Delta/\kB T)$ between the molecular Einstein coefficients for absorption $B_\mathrm{abs}$ and emission $B_\mathrm{em}$ of condensate photons. This detailed balance condition for large negative $\Delta$ energetically favors a relatively large photon number $\langle n\rangle $ as compared to the (generally larger) number of excited molecules $\langle M_\mathrm{e}\rangle$, see the left panel in Fig.~\ref{fig:1}(b); for $\Delta\rightarrow 0^{-}$, on the other hand, $\langle M_\mathrm{e}\rangle$ by far exceeds $\langle n\rangle$ and presents the dominant contribution to $X$ (right panel). Correspondingly, we can define an effective reservoir size $M_\mathrm{eff}=M/\left[2+2 \cosh(\hbar\Delta/\kB T)\right]$ based on the condition $\mathcal{P}_0=\mathcal{P}_1$~\cite{Klaers:2012,Schmitt:2014}. At this point, which occurs for $\langle n\rangle^2=M_\mathrm{eff}$, the photon number statistics changes from super-Poissonian to Poissonian, realizing a distinction point between the grand canonical and canonical statistical regimes~\cite{Schmitt:2018}.

Using eq.~\eqref{eq:1}, the lowest moments $\langle n^k\rangle=\sum_n n^k \mathcal{P}_n$ can be expressed through first and second-order derivatives of the partition function, $\langle n\rangle= -Z^{-1}(\kB T/\hbar) \frac{dZ}{d\Delta}$ and $\langle n^2\rangle = Z^{-1} (\kB T/\hbar)^2 \frac{d^2 Z}{d\Delta^2}$. We obtain an expression forming a fluctuation-dissipation relation
\begin{equation}
    \langle \Delta n^2\rangle=-\frac{\kB T}{\hbar} \left(\frac{d \langle{n}\rangle}{d\Delta}\right)_{X,T},
\label{eq:3}
\end{equation}
which connects the squared photon number fluctuations $\langle \Delta n^2\rangle = \langle n^2\rangle - \langle{n}\rangle^2$ to a reactive response function $(d \langle{n}\rangle/d\Delta)_{X,T}$ by thermal energy $\kB T$ (in angular frequency units). The response function shown in the bottom panel of Fig.~\ref{fig:1}(b) describes the susceptibility of the mean condensate population $\langle{n}\rangle$ to changes of the dye-cavity detuning $\Delta$ at constant temperature $T$ and excitation number $X$; in other words, it qualifies how easy it is to "compress" photons into the dye reservoir. The fluctuation-dissipation relation in eq.~\eqref{eq:3} directly translates (intrinsic) thermal energy fluctuations of the dye molecules determined by $\kB T$ into the magnitude of particle number fluctuations via a scaling factor determined by the response function ${\hbar^{-1}}\left(d \langle{n}\rangle/d\Delta\right)_{X,T}$.

Intuitively, the sharp increase of the response visible in Fig.~\ref{fig:1}(b) can be understood from an energy argument: At each detuning $\Delta$, the system minimizes its free energy $F=E-TS$ by balancing the energy reduction $E=n\hbar\Delta$ from creating photons and the entropy $S=\kB \ln\{M!/[M_\mathrm{e}!(M-M_\mathrm{e})!]\}$ in the dye molecules. At sufficiently negative detunings, the energy gained by creating photons prevails and $F$ becomes minimal for $\langle n\rangle = X- M/[1+\exp(-\hbar\Delta/\kB T)]$. The onset of this photon production occurs at $\hbar\Delta/\kB T = \ln[X/(M-X)]$ and exhibits a steep slope for large excitation numbers.

%%%%%%%%%%%%%%%%%
%%%%% FIG 2 %%%%%
%%%%%%%%%%%%%%%%%
\begin{figure}[t]
    \centering
    \includegraphics[width=1.0\columnwidth]{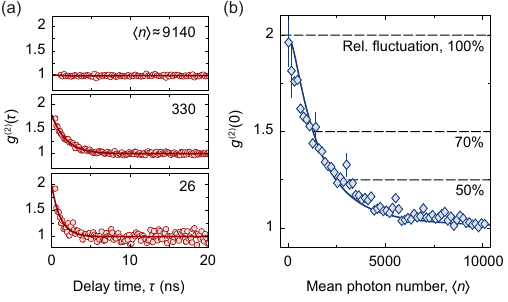}
    \caption{Number fluctuations in the condensate mode from the canonical to the grand canonical ensemble for a fixed dye-cavity detuning $\Delta = -4.571\kB T/\hbar$. (a) Second-order correlation functions $g^{(2)}(\tau)$ for decreasing photon numbers $\langle n\rangle$ along with fits (solid). For the largest $\langle n\rangle\approx 9\thinspace 140$ (top panel), the condensate is second-order coherent with $g^{(2)}(0) = 1.02\pm 0.08$, indicating Poissonian number statistics under canonical conditions. For smaller $\langle n\rangle\approx \{330,26\}$, photon bunching in the grand canonical condensate emission is observed with $g^{(2)}(0) =\{ 1.79\pm 0.20,1.96\pm 0.09\}$. (b) The zero-delay $g^{(2)}(0)$ as a function of $\langle n\rangle$ shows the crossover of the number statistics from the grand canonical regime with $\sqrt{\langle \Delta n^2\rangle}/\langle n\rangle=1$ to the canonical one with $\sqrt{\langle \Delta n^2\rangle }/\langle n\rangle=0$, along with theory for $M=1.8\cdot 10^8$ (solid line). Fluctuation levels are indicated by dashed lines, and error bars are calculated from the uncertainties of the fit parameters. The fit function $f(\tau)=1+C_1 e^{\lambda_1 \tau}+C_2 e^{\lambda_2 \tau}$~\cite{Oeztuerk:2021}, where $\lambda_{1,2}=-\delta\pm\sqrt{\delta^2-\omega_0^2}$ and $C_{1,2}=Y\pm i\sqrt{\omega_0^2-\delta^2}Z$, gives the fit parameters $\delta\approx\{0.38, 0.42, 0.45\}~\SI{}{\nano\second}^{-1}$, $\omega_0\approx\{0.43, 0.41, 0.10\}~\SI{}{\nano\second}^{-1}$, $Y\approx\{0.01, 0.39, 0.48\}$, $Z\approx\{0.15, 5.63, 1.09\}$ for the three cases in (a).}
    \label{fig:2}
\end{figure}

To measure the number statistics, fluctuations and response function of the photon condensate, the cavity emission is dispersed on a high-resolution grating, and the spectrally filtered condensate is recorded by time-resolved photon counting on a streak camera, see Fig.~\ref{fig:1}(a). The grating resolution $\SI{18}{\giga\hertz}<\Omega/(2\pi)$ allows us to not only accurately determine $\Delta$, but also to separate the condensate photons from the thermal cloud and measure genuine single-mode correlations without admixing of uncorrelated photons in the lowest-lying excited modes; the time-integrated spectrum in Fig.~\ref{fig:1}(a) shows the clear separation of the condensate signal from higher-mode contributions. Figure~\ref{fig:1}(c) shows exemplary data of the (attenuated) time-resolved condensate emission during a single streak camera exposure, from which both the photon number statistics and the second-order correlations are derived after averaging over many such traces. While for large negative detunings photons are randomly distributed in time (top panel), a condensate emitting bunches of photons is observed for detunings closer to resonance
(bottom). From our measurements, we extract the condensate population and its fluctuation for different detunings $\Delta$, realized by changing the cavity length. The average condensate number $\langle n\rangle$ is determined by monitoring part of the condensate emission on a calibrated photomultiplier, while the fluctuations are inferred from the second-order correlations, providing the required input to validate eq.~\eqref{eq:3}.

Figure~\ref{fig:2}(a) shows measured second-order correlations $g^{(2)}(\tau)$ as a function of the delay time $\tau$, see Fig.~\ref{fig:1}(c), for different average photon numbers $\langle n\rangle$ along with fits (lines)~\cite{Oeztuerk:2021} for fixed detuning. For large times, $g^{(2)}(\tau)$ generically decays to unity, but for $\tau\approx 0$ and moderate $\langle n\rangle$, photon bunching is observed. Figure~\ref{fig:2}(b) shows the variation of the fitted zero-delay correlation $g^{(2)}(0) =\langle n^2\rangle/\langle n\rangle^2$ versus $\langle n\rangle$ in good agreement with theory (line). Notably, this bunching amplitude provides a direct measure for the number fluctuations $\langle\Delta n^2\rangle = (g^{(2)}(0)-1)\langle n\rangle^2$, as contained in eq.~\eqref{eq:3}. For a condensate population with $\langle n\rangle < \sqrt{M_\mathrm{eff}}$, the observed bunching indicates the grand canonical statistical regime. When increasing the photon number, $g^{(2)}(0)$ gradually approaches unity, signalling the number statistics of a “usual” condensate in the canonical statistical ensemble. For the data in Fig.~\ref{fig:2}, we have $\sqrt{M_\mathrm{eff}}\approx 1350$. For the smallest investigated condensate mode occupation with $\langle n \rangle= 26$, we find $g^{(2)}(0)\simeq1.96(9)$, which within experimental uncertainties agrees with the grand canonical prediction $\langle\Delta n^2\rangle/\langle n\rangle^2=1$.

\begin{figure}[t]
    \centering
    \includegraphics[width=1.0\columnwidth]{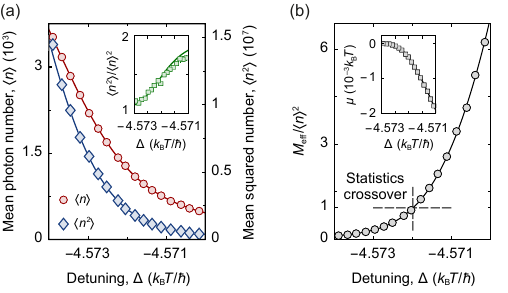}
    \caption{
    (a) First and second moment of the photon number, $\langle n\rangle$ (red circles) and $\langle n^2\rangle$ (blue diamonds), respectively, as a function of the detuning $\Delta$ for $X\approx 1.538\cdot 10^6$ and $M=1.5\cdot 10^8$ along with theory (solid). As the modulus of the detuning $|\Delta|$ is reduced, the condensate population $\langle n\rangle$ decreases, as here more system excitations are present in the form of excited molecules. The second moment $\langle n^2\rangle$ decays with a different slope, which is more clearly visible when comparing it to $\langle n\rangle^2$, showing the increase of $g^{(2)}(0)=\langle n^2\rangle/\langle n\rangle^2$ (green squares, inset); at $\Delta=-4.572\kB T/\hbar$ the statistics crosses over from canonical to grand canonical conditions. (b) Relative effective reservoir size $M_\mathrm{eff}/\langle n\rangle^2$ and chemical potential $\mu$ (inset) versus detuning. The particle reservoir grows to $M_\mathrm{eff} > 6 \langle n\rangle^2$ for small $|\Delta|$ and to good approximation establishes grand canonical conditions (with $\mu< 0$). In the opposite limit of large negative $\Delta$, the particle reservoir shrinks and canonical conditions with $\mu\approx0$ apply.}
    \label{fig:3}
\end{figure}

\begin{figure}[t]
    \centering
    \includegraphics[width=1.0\columnwidth]{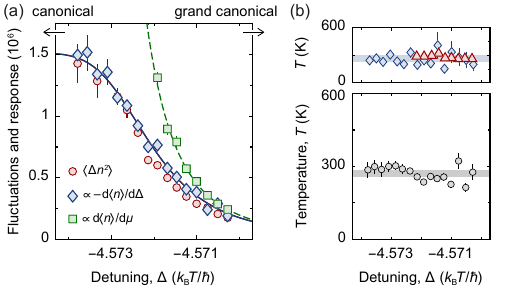}
    \caption{(a) Fluctuation-dissipation relation in a room temperature Bose-Einstein condensate. The squared number fluctuations $\langle \Delta n^2\rangle$ (red circles) and the scaled response function $-\kB T/\hbar (d\langle n\rangle/d\Delta)_{X,T}$ (blue diamonds) as a function of the detuning $\Delta$ resemble both sides of eq.~\eqref{eq:3}. For comparison, we also show the scaled compressibility $\kB T (d\langle n\rangle/d\mu)_{T}$ (green squares), which agrees with $\langle \Delta n^2\rangle$ only in the grand canonical limit due to the here non-vanishing chemical potential. Lines give theory based on eq.~\eqref{eq:1}. (b) The spectral temperature of the photon condensate as obtained from the ratio of the independently measured $\langle \Delta n^2\rangle$ and $-\kB/\hbar (d\langle n\rangle/d\Delta)_{X,T}$ averages to $T=271(30)~\SI{}{\kelvin}$, which within its standard deviation (shading) agrees with the ambient temperature $\approx \SI{300}{\kelvin}$. The top panel shows extracted temperatures for other target values of $X\approx\{1.535,1.532\}\cdot 10^6$ (blue diamonds, red triangles).}
    \label{fig:4}
\end{figure}

To determine the reactive response function of the condensate, see the right-hand side of eq.~\eqref{eq:3}, requires the derivative $(d\langle n\rangle / d\Delta)_{X,T}$ to be evaluated at constant temperature $T$ and excitation number $X=n+M_\mathrm{e}$. The molecular part $M_\mathrm{e}$ is not directly accessible in our experiments, we must therefore determine $X$ indirectly from the precisely measured $\langle n\rangle$ and $\Delta$. We reconstruct $X$ by fitting $\langle n\rangle$, recorded at different pump powers and dye-cavity detunings, with theory $\sum_n {n\mathcal{P}_{n}}$ based on eq.~\eqref{eq:1}; here we use only $X$ as a fit parameter, while the molecule number $M$ and $\Delta$ are fixed. Each measurement thus yields four-fold information $\{g^{(2)}(0),\langle n\rangle,\Delta,X\}$, from which only the data closest to the target value $X\approx 1.538\cdot 10^6$ are retained for further analysis. Note that we have also examined other target values of $X$, but find better statistics at the selected one (see the top panel in Fig.~\ref{fig:4}(b)). Figure~\ref{fig:3}(a) shows the obtained first and second moment of the photon number as a function of the detuning for the corresponding, fixed excitation number. For large negative detunings, strongly occupied condensates with suppressed fluctuations indicate the canonical statistical regime. In the opposite limit of detunings closer to resonance, a large fraction of $X$ consists of dye electronic excitations, forming the particle reservoir; here the reduced condensate occupancy with significant fluctuations signals the onset of grand canonical statistics. The interesting range of detunings, where the fluctuations are significantly varying for constant $X$ (Fig.~\ref{fig:3}(a), inset), is spectrally narrow and covers only $0.004 \kB T/\hbar$ ($\approx \SI{25}{\giga\hertz}$), as well understood from the large number of excitations $X$ stored in the system.

%%%%%%%%%%%%%%%%%
%%%%% FIG 3 %%%%%
%%%%%%%%%%%%%%%%%

Figure~\ref{fig:3}(b) shows the behavior of the relative effective reservoir size $M_\mathrm{eff}/\langle n\rangle^2$, which takes values below to well above unity as the detuning is tuned closer to resonance. The inset of Fig.~\ref{fig:3}(b) gives the photon chemical potential $\mu = \kB T \ln\left[ (X-\langle n\rangle)/(M-X+\langle n\rangle)\right]-\hbar\Delta$, which directly depends on $X$ via the ratio between excited and ground state molecules~\cite{Klaers:2012,Schmitt:2018}. At large negative detunings, $\mu\simeq 0$, as expected from Bose-Einstein statistics. Notably, in the opposite, grand canonical limit, the chemical potential becomes finite; nevertheless, $|\mu|< \hbar\Omega$ indicates that despite fluctuations the system remains condensed.

%%%%%%%%%%%%%%%%%
%%%%% FIG 4 %%%%%
%%%%%%%%%%%%%%%%%

Next we examine the validity of the fluctuation-dissipation relation in eq.~\eqref{eq:3} by directly comparing the condensate number fluctuations (see Fig.~\ref{fig:3}) to the response function for varied detuning. Figure~\ref{fig:4}(a) shows the resulting squared number fluctuations $\langle \Delta n^2\rangle$ (red circles) and the scaled response function $-\kB T/\hbar (d\langle n\rangle /d\Delta)_{X,T}$ (blue diamonds) as a function of the detuning $\Delta$, here for fixed $T=\SI{300}{\kelvin}$. The good agreement between both data sets, and with theory (solid line), gives evidence for the fluctuation-dissipation relation to be well fulfilled in our system both in the canonical and grand canonical regime. We note that in the latter case the fluctuation-dissipation relation can be written in terms of the isothermal compressibility $\kappa_T=(d\langle n\rangle / d\mu)_{X,T}$~\cite{Busley:2022}. We use the derivative $\left( d\mu/d\langle n \rangle\right)_{X,T}\simeq -\hbar\left( d\Delta / d\langle n\rangle \right)_{X,T} - \kB T / (2 M_\mathrm{eff})$ to rewrite eq.~\eqref{eq:3} and obtain 
\begin{equation}
    \langle \Delta n^2\rangle = \frac{1}{\frac{1}{2M_\mathrm{eff}} + \frac{1}{\kB T \left(\frac{d\langle n\rangle}{d\mu}\right)_{X,T}}} \stackrel{M_\mathrm{eff}\rightarrow\infty}{\longrightarrow} \kB T \kappa_T,
\label{eq:4}
\end{equation}
which for large reservoirs $M_\mathrm{eff}$, corresponding to the grand canonical regime, approaches the "textbook" form of the fluctuation-dissipation relation~\cite{Huang:1987}. The data (green squares) in Fig.~\ref{fig:4}(a) show that the correction term $1/(2 M_\mathrm{eff})$ can be neglected only deep in the grand canonical regime, while for large negative detunings the compressibility diverges.

Our measurements confirm the thermal nature of the room-temperature Bose-Einstein condensate of photons in a rigorous way. Figure~\ref{fig:4}(b) shows the deduced temperature $T = -{\hbar/\kB \langle \Delta n^2\rangle} / {\left(d\langle n\rangle / d\Delta\right)_{X,T}}$ as a function of the dye-cavity detuning, which agrees with room temperature over the investigated range of detunings, and we find $T=271(30)~\SI{}{\kelvin}$. Physically, the results give evidence that independent of the detuning the statistical number fluctuations are driven by thermal energy, with correspondingly varying “stiffness” of the response to perturbations.

In conclusion, we have demonstrated a fluctuation-dissipation relation connecting the statistical number fluctuations of a photon Bose-Einstein condensate coupled to a dye reservoir with the reactive response of the condensate population to variations of the relative energy scale of system and reservoir constituents. The ratio between the independently measured fluctuations and response function agrees with thermal energy, confirming the thermalized nature of the optical condensate. In the same breath, the findings also give evidence for the thermal equilibrium character of the molecular reservoir. For the future, probing the dynamical response of the system after a fast perturbation of the reservoir can allow us to extend the studies of the fluctuation-dissipation relation to the time-dependent regime. Other prospects include studies of fluctuations and susceptibilities associated with photon transport, {\it e.g.}, in lattice or box geometries~\cite{Gladilin:2020a,Busley:2022}, and their applicability to open quantum systems, which become accessible when tuning thermalization and photon loss~\cite{Oeztuerk:2021}.

We thank I. Carusotto, W. Ketterle, and W. D. Phillips for discussions, and P. Christodoulou, J. Anglin and D. Luitz for comments on the manuscript. We acknowledge funding by the DFG within SFB/TR 185 (277625399) and the Cluster of Excellence ML4Q (EXC 2004/1–390534769), the EU within the Quantum Flagship project PhoQuS (Grant No. 820392), and the DLR with funds provided by the BMWi (50WM1859). J.S. acknowledges support from an ML4Q Independence grant.

\bibliography{references}

\begin{thebibliography}{41}%
\makeatletter
\providecommand \@ifxundefined [1]{%
 \@ifx{#1\undefined}
}%
\providecommand \@ifnum [1]{%
 \ifnum #1\expandafter \@firstoftwo
 \else \expandafter \@secondoftwo
 \fi
}%
\providecommand \@ifx [1]{%
 \ifx #1\expandafter \@firstoftwo
 \else \expandafter \@secondoftwo
 \fi
}%
\providecommand \natexlab [1]{#1}%
\providecommand \enquote  [1]{``#1''}%
\providecommand \bibnamefont  [1]{#1}%
\providecommand \bibfnamefont [1]{#1}%
\providecommand \citenamefont [1]{#1}%
\providecommand \href@noop [0]{\@secondoftwo}%
\providecommand \href [0]{\begingroup \@sanitize@url \@href}%
\providecommand \@href[1]{\@@startlink{#1}\@@href}%
\providecommand \@@href[1]{\endgroup#1\@@endlink}%
\providecommand \@sanitize@url [0]{\catcode `\\12\catcode `\$12\catcode
  `\&12\catcode `\#12\catcode `\^12\catcode `\_12\catcode `\%12\relax}%
\providecommand \@@startlink[1]{}%
\providecommand \@@endlink[0]{}%
\providecommand \url  [0]{\begingroup\@sanitize@url \@url }%
\providecommand \@url [1]{\endgroup\@href {#1}{\urlprefix }}%
\providecommand \urlprefix  [0]{URL }%
\providecommand \Eprint [0]{\href }%
\providecommand \doibase [0]{https://doi.org/}%
\providecommand \selectlanguage [0]{\@gobble}%
\providecommand \bibinfo  [0]{\@secondoftwo}%
\providecommand \bibfield  [0]{\@secondoftwo}%
\providecommand \translation [1]{[#1]}%
\providecommand \BibitemOpen [0]{}%
\providecommand \bibitemStop [0]{}%
\providecommand \bibitemNoStop [0]{.\EOS\space}%
\providecommand \EOS [0]{\spacefactor3000\relax}%
\providecommand \BibitemShut  [1]{\csname bibitem#1\endcsname}%
\let\auto@bib@innerbib\@empty
%</preamble>
\bibitem [{\citenamefont {Callen}\ and\ \citenamefont
  {Welton}(1951)}]{Callen:1951}%
  \BibitemOpen
  \bibfield  {author} {\bibinfo {author} {\bibfnamefont {H.~B.}\ \bibnamefont
  {Callen}}\ and\ \bibinfo {author} {\bibfnamefont {T.~A.}\ \bibnamefont
  {Welton}},\ }\bibfield  {title} {\bibinfo {title} {Irreversibility and
  generalized noise},\ }\href {https://doi.org/10.1103/PhysRev.83.34}
  {\bibfield  {journal} {\bibinfo  {journal} {Phys. Rev.}\ }\textbf {\bibinfo
  {volume} {83}},\ \bibinfo {pages} {34} (\bibinfo {year} {1951})}\BibitemShut
  {NoStop}%
\bibitem [{\citenamefont {Kubo}(1966)}]{Kubo:1966}%
  \BibitemOpen
  \bibfield  {author} {\bibinfo {author} {\bibfnamefont {R.}~\bibnamefont
  {Kubo}},\ }\bibfield  {title} {\bibinfo {title} {The fluctuation-dissipation
  theorem},\ }\href {https://doi.org/10.1088/0034-4885/29/1/306} {\bibfield
  {journal} {\bibinfo  {journal} {Rep. Prog. Phys.}\ }\textbf {\bibinfo
  {volume} {29}},\ \bibinfo {pages} {255} (\bibinfo {year} {1966})}\BibitemShut
  {NoStop}%
\bibitem [{\citenamefont {Einstein}(1905)}]{Einstein:1905}%
  \BibitemOpen
  \bibfield  {author} {\bibinfo {author} {\bibfnamefont {A.}~\bibnamefont
  {Einstein}},\ }\bibfield  {title} {\bibinfo {title} {{\"U}ber die von der
  molekularkinetischen {T}heorie der {W}\"arme geforderte {B}ewegung von in
  ruhenden {F}l\"ussigkeiten suspendierten {T}eilchen},\ }\href
  {https://doi.org/10.1002/andp.19053220806} {\bibfield  {journal} {\bibinfo
  {journal} {Ann. d. Phys. (Leipzig)}\ }\textbf {\bibinfo {volume} {17}},\
  \bibinfo {pages} {549} (\bibinfo {year} {1905})}\BibitemShut {NoStop}%
\bibitem [{\citenamefont {Johnson}(1928)}]{Johnson:1928}%
  \BibitemOpen
  \bibfield  {author} {\bibinfo {author} {\bibfnamefont {J.~B.}\ \bibnamefont
  {Johnson}},\ }\bibfield  {title} {\bibinfo {title} {Thermal agitation of
  electricity in conductors},\ }\href {https://doi.org/10.1103/PhysRev.32.97}
  {\bibfield  {journal} {\bibinfo  {journal} {Phys. Rev.}\ }\textbf {\bibinfo
  {volume} {32}},\ \bibinfo {pages} {97} (\bibinfo {year} {1928})}\BibitemShut
  {NoStop}%
\bibitem [{\citenamefont {Esteve}\ \emph {et~al.}(2006)\citenamefont {Esteve},
  \citenamefont {Trebbia}, \citenamefont {Schumm}, \citenamefont {Aspect},
  \citenamefont {Westbrook},\ and\ \citenamefont {Bouchoule}}]{Esteve:2006}%
  \BibitemOpen
  \bibfield  {author} {\bibinfo {author} {\bibfnamefont {J.}~\bibnamefont
  {Esteve}}, \bibinfo {author} {\bibfnamefont {J.-B.}\ \bibnamefont {Trebbia}},
  \bibinfo {author} {\bibfnamefont {T.}~\bibnamefont {Schumm}}, \bibinfo
  {author} {\bibfnamefont {A.}~\bibnamefont {Aspect}}, \bibinfo {author}
  {\bibfnamefont {C.~I.}\ \bibnamefont {Westbrook}},\ and\ \bibinfo {author}
  {\bibfnamefont {I.}~\bibnamefont {Bouchoule}},\ }\bibfield  {title} {\bibinfo
  {title} {Observations of density fluctuations in an elongated {B}ose gas:
  {I}deal gas and quasicondensate regimes},\ }\href
  {https://doi.org/10.1103/PhysRevLett.96.130403} {\bibfield  {journal}
  {\bibinfo  {journal} {Phys. Rev. Lett.}\ }\textbf {\bibinfo {volume} {96}},\
  \bibinfo {pages} {130403} (\bibinfo {year} {2006})}\BibitemShut {NoStop}%
\bibitem [{\citenamefont {M\"uller}\ \emph {et~al.}(2010)\citenamefont
  {M\"uller}, \citenamefont {Zimmermann}, \citenamefont {Meineke},
  \citenamefont {Brantut}, \citenamefont {Esslinger},\ and\ \citenamefont
  {Moritz}}]{Mueller:2010}%
  \BibitemOpen
  \bibfield  {author} {\bibinfo {author} {\bibfnamefont {T.}~\bibnamefont
  {M\"uller}}, \bibinfo {author} {\bibfnamefont {B.}~\bibnamefont
  {Zimmermann}}, \bibinfo {author} {\bibfnamefont {J.}~\bibnamefont {Meineke}},
  \bibinfo {author} {\bibfnamefont {J.-P.}\ \bibnamefont {Brantut}}, \bibinfo
  {author} {\bibfnamefont {T.}~\bibnamefont {Esslinger}},\ and\ \bibinfo
  {author} {\bibfnamefont {H.}~\bibnamefont {Moritz}},\ }\bibfield  {title}
  {\bibinfo {title} {Local observation of antibunching in a trapped {F}ermi
  gas},\ }\href {https://doi.org/10.1103/PhysRevLett.105.040401} {\bibfield
  {journal} {\bibinfo  {journal} {Phys. Rev. Lett.}\ }\textbf {\bibinfo
  {volume} {105}},\ \bibinfo {pages} {040401} (\bibinfo {year}
  {2010})}\BibitemShut {NoStop}%
\bibitem [{\citenamefont {Sanner}\ \emph {et~al.}(2010)\citenamefont {Sanner},
  \citenamefont {Su}, \citenamefont {Keshet}, \citenamefont {Gommers},
  \citenamefont {Shin}, \citenamefont {Huang},\ and\ \citenamefont
  {Ketterle}}]{Sanner:2010}%
  \BibitemOpen
  \bibfield  {author} {\bibinfo {author} {\bibfnamefont {C.}~\bibnamefont
  {Sanner}}, \bibinfo {author} {\bibfnamefont {E.~J.}\ \bibnamefont {Su}},
  \bibinfo {author} {\bibfnamefont {A.}~\bibnamefont {Keshet}}, \bibinfo
  {author} {\bibfnamefont {R.}~\bibnamefont {Gommers}}, \bibinfo {author}
  {\bibfnamefont {Y.-i.}\ \bibnamefont {Shin}}, \bibinfo {author}
  {\bibfnamefont {W.}~\bibnamefont {Huang}},\ and\ \bibinfo {author}
  {\bibfnamefont {W.}~\bibnamefont {Ketterle}},\ }\bibfield  {title} {\bibinfo
  {title} {Suppression of density fluctuations in a quantum degenerate {F}ermi
  gas},\ }\href {https://doi.org/10.1103/PhysRevLett.105.040402} {\bibfield
  {journal} {\bibinfo  {journal} {Phys. Rev. Lett.}\ }\textbf {\bibinfo
  {volume} {105}},\ \bibinfo {pages} {040402} (\bibinfo {year}
  {2010})}\BibitemShut {NoStop}%
\bibitem [{\citenamefont {Gemelke}\ \emph {et~al.}(2009)\citenamefont
  {Gemelke}, \citenamefont {Zhang}, \citenamefont {Hung},\ and\ \citenamefont
  {Chin}}]{Gemelke:2009}%
  \BibitemOpen
  \bibfield  {author} {\bibinfo {author} {\bibfnamefont {N.}~\bibnamefont
  {Gemelke}}, \bibinfo {author} {\bibfnamefont {X.}~\bibnamefont {Zhang}},
  \bibinfo {author} {\bibfnamefont {C.-L.}\ \bibnamefont {Hung}},\ and\
  \bibinfo {author} {\bibfnamefont {C.}~\bibnamefont {Chin}},\ }\bibfield
  {title} {\bibinfo {title} {In situ observation of incompressible
  {M}ott-insulating domains in ultracold atomic gases},\ }\href
  {https://doi.org/10.1038/nature08244} {\bibfield  {journal} {\bibinfo
  {journal} {Nature}\ }\textbf {\bibinfo {volume} {460}},\ \bibinfo {pages}
  {995} (\bibinfo {year} {2009})}\BibitemShut {NoStop}%
\bibitem [{\citenamefont {Hung}\ \emph {et~al.}(2011)\citenamefont {Hung},
  \citenamefont {Zhang}, \citenamefont {Gemelke},\ and\ \citenamefont
  {Chin}}]{Hung:2011}%
  \BibitemOpen
  \bibfield  {author} {\bibinfo {author} {\bibfnamefont {C.-L.}\ \bibnamefont
  {Hung}}, \bibinfo {author} {\bibfnamefont {X.}~\bibnamefont {Zhang}},
  \bibinfo {author} {\bibfnamefont {N.}~\bibnamefont {Gemelke}},\ and\ \bibinfo
  {author} {\bibfnamefont {C.}~\bibnamefont {Chin}},\ }\bibfield  {title}
  {\bibinfo {title} {Observation of scale invariance and universality in
  two-dimensional {B}ose gases},\ }\href {https://doi.org/10.1038/nature09722}
  {\bibfield  {journal} {\bibinfo  {journal} {Nature}\ }\textbf {\bibinfo
  {volume} {470}},\ \bibinfo {pages} {236} (\bibinfo {year}
  {2011})}\BibitemShut {NoStop}%
\bibitem [{\citenamefont {Blumkin}\ \emph {et~al.}(2013)\citenamefont
  {Blumkin}, \citenamefont {Rinott}, \citenamefont {Schley}, \citenamefont
  {Berkovitz}, \citenamefont {Shammass},\ and\ \citenamefont
  {Steinhauer}}]{Blumkin:2013}%
  \BibitemOpen
  \bibfield  {author} {\bibinfo {author} {\bibfnamefont {A.}~\bibnamefont
  {Blumkin}}, \bibinfo {author} {\bibfnamefont {S.}~\bibnamefont {Rinott}},
  \bibinfo {author} {\bibfnamefont {R.}~\bibnamefont {Schley}}, \bibinfo
  {author} {\bibfnamefont {A.}~\bibnamefont {Berkovitz}}, \bibinfo {author}
  {\bibfnamefont {I.}~\bibnamefont {Shammass}},\ and\ \bibinfo {author}
  {\bibfnamefont {J.}~\bibnamefont {Steinhauer}},\ }\bibfield  {title}
  {\bibinfo {title} {Observing atom bunching by the fourier slice theorem},\
  }\href {https://doi.org/10.1103/PhysRevLett.110.265301} {\bibfield  {journal}
  {\bibinfo  {journal} {Phys. Rev. Lett.}\ }\textbf {\bibinfo {volume} {110}},\
  \bibinfo {pages} {265301} (\bibinfo {year} {2013})}\BibitemShut {NoStop}%
\bibitem [{\citenamefont {Christodoulou}\ \emph {et~al.}(2021)\citenamefont
  {Christodoulou}, \citenamefont {Gałka}, \citenamefont {Dogra}, \citenamefont
  {Lopes}, \citenamefont {Schmitt},\ and\ \citenamefont
  {Hadzibabic}}]{Christodoulou:2021}%
  \BibitemOpen
  \bibfield  {author} {\bibinfo {author} {\bibfnamefont {P.}~\bibnamefont
  {Christodoulou}}, \bibinfo {author} {\bibfnamefont {M.}~\bibnamefont
  {Gałka}}, \bibinfo {author} {\bibfnamefont {N.}~\bibnamefont {Dogra}},
  \bibinfo {author} {\bibfnamefont {R.}~\bibnamefont {Lopes}}, \bibinfo
  {author} {\bibfnamefont {J.}~\bibnamefont {Schmitt}},\ and\ \bibinfo {author}
  {\bibfnamefont {Z.}~\bibnamefont {Hadzibabic}},\ }\bibfield  {title}
  {\bibinfo {title} {Observation of first and second sound in a {BKT}
  superfluid},\ }\href {https://doi.org/10.1038/s41586-021-03537-9} {\bibfield
  {journal} {\bibinfo  {journal} {Nature}\ }\textbf {\bibinfo {volume} {594}},\
  \bibinfo {pages} {191} (\bibinfo {year} {2021})}\BibitemShut {NoStop}%
\bibitem [{\citenamefont {Kristensen}\ \emph {et~al.}(2019)\citenamefont
  {Kristensen}, \citenamefont {Christensen}, \citenamefont {Gajdacz},
  \citenamefont {Iglicki}, \citenamefont {Paw\l{}owski}, \citenamefont
  {Klempt}, \citenamefont {Sherson}, \citenamefont {Rza\ifmmode~\dot{z}\else
  \.{z}\fi{}ewski}, \citenamefont {Hilliard},\ and\ \citenamefont
  {Arlt}}]{Kristensen:2019}%
  \BibitemOpen
  \bibfield  {author} {\bibinfo {author} {\bibfnamefont {M.~A.}\ \bibnamefont
  {Kristensen}}, \bibinfo {author} {\bibfnamefont {M.~B.}\ \bibnamefont
  {Christensen}}, \bibinfo {author} {\bibfnamefont {M.}~\bibnamefont
  {Gajdacz}}, \bibinfo {author} {\bibfnamefont {M.}~\bibnamefont {Iglicki}},
  \bibinfo {author} {\bibfnamefont {K.}~\bibnamefont {Paw\l{}owski}}, \bibinfo
  {author} {\bibfnamefont {C.}~\bibnamefont {Klempt}}, \bibinfo {author}
  {\bibfnamefont {J.~F.}\ \bibnamefont {Sherson}}, \bibinfo {author}
  {\bibfnamefont {K.}~\bibnamefont {Rza\ifmmode~\dot{z}\else \.{z}\fi{}ewski}},
  \bibinfo {author} {\bibfnamefont {A.~J.}\ \bibnamefont {Hilliard}},\ and\
  \bibinfo {author} {\bibfnamefont {J.~J.}\ \bibnamefont {Arlt}},\ }\bibfield
  {title} {\bibinfo {title} {Observation of atom number fluctuations in a
  {B}ose-{E}instein condensate},\ }\href
  {https://doi.org/10.1103/PhysRevLett.122.163601} {\bibfield  {journal}
  {\bibinfo  {journal} {Phys. Rev. Lett.}\ }\textbf {\bibinfo {volume} {122}},\
  \bibinfo {pages} {163601} (\bibinfo {year} {2019})}\BibitemShut {NoStop}%
\bibitem [{\citenamefont {Christensen}\ \emph {et~al.}(2021)\citenamefont
  {Christensen}, \citenamefont {Vibel}, \citenamefont {Hilliard}, \citenamefont
  {Kruk}, \citenamefont {Paw\l{}owski}, \citenamefont {Hryniuk}, \citenamefont
  {Rza\ifmmode \mbox{\c{}}\else \c{}\fi{}\ifmmode~\dot{z}\else
  \.{z}\fi{}ewski}, \citenamefont {Kristensen},\ and\ \citenamefont
  {Arlt}}]{Christensen:2021}%
  \BibitemOpen
  \bibfield  {author} {\bibinfo {author} {\bibfnamefont {M.~B.}\ \bibnamefont
  {Christensen}}, \bibinfo {author} {\bibfnamefont {T.}~\bibnamefont {Vibel}},
  \bibinfo {author} {\bibfnamefont {A.~J.}\ \bibnamefont {Hilliard}}, \bibinfo
  {author} {\bibfnamefont {M.~B.}\ \bibnamefont {Kruk}}, \bibinfo {author}
  {\bibfnamefont {K.}~\bibnamefont {Paw\l{}owski}}, \bibinfo {author}
  {\bibfnamefont {D.}~\bibnamefont {Hryniuk}}, \bibinfo {author} {\bibfnamefont
  {K.}~\bibnamefont {Rza\ifmmode \mbox{\c{}}\else
  \c{}\fi{}\ifmmode~\dot{z}\else \.{z}\fi{}ewski}}, \bibinfo {author}
  {\bibfnamefont {M.~A.}\ \bibnamefont {Kristensen}},\ and\ \bibinfo {author}
  {\bibfnamefont {J.~J.}\ \bibnamefont {Arlt}},\ }\bibfield  {title} {\bibinfo
  {title} {Observation of microcanonical atom number fluctuations in a
  {Bose}-{E}instein condensate},\ }\href
  {https://doi.org/10.1103/PhysRevLett.126.153601} {\bibfield  {journal}
  {\bibinfo  {journal} {Phys. Rev. Lett.}\ }\textbf {\bibinfo {volume} {126}},\
  \bibinfo {pages} {153601} (\bibinfo {year} {2021})}\BibitemShut {NoStop}%
\bibitem [{\citenamefont {Bloch}\ \emph {et~al.}(2022)\citenamefont {Bloch},
  \citenamefont {Carusotto},\ and\ \citenamefont {Wouters}}]{Bloch:2022}%
  \BibitemOpen
  \bibfield  {author} {\bibinfo {author} {\bibfnamefont {J.}~\bibnamefont
  {Bloch}}, \bibinfo {author} {\bibfnamefont {I.}~\bibnamefont {Carusotto}},\
  and\ \bibinfo {author} {\bibfnamefont {M.}~\bibnamefont {Wouters}},\
  }\bibfield  {title} {\bibinfo {title} {Non-equilibrium {B}ose--{E}instein
  condensation in photonic systems},\ }\href
  {https://doi.org/10.1038/s42254-022-00464-0} {\bibfield  {journal} {\bibinfo
  {journal} {Nat. Rev. Phys.}\ }\textbf {\bibinfo {volume} {4}},\ \bibinfo
  {pages} {470–488} (\bibinfo {year} {2022})}\BibitemShut {NoStop}%
\bibitem [{\citenamefont {Agarwal}(1972)}]{Agarwal:1972}%
  \BibitemOpen
  \bibfield  {author} {\bibinfo {author} {\bibfnamefont {G.}~\bibnamefont
  {Agarwal}},\ }\bibfield  {title} {\bibinfo {title} {Fluctuation-dissipation
  theorems for systems in non-thermal equilibrium and application to laser
  light},\ }\href
  {https://doi.org/https://doi.org/10.1016/0375-9601(72)90502-6} {\bibfield
  {journal} {\bibinfo  {journal} {Phys. Lett. A}\ }\textbf {\bibinfo {volume}
  {38}},\ \bibinfo {pages} {93} (\bibinfo {year} {1972})}\BibitemShut {NoStop}%
\bibitem [{\citenamefont {H{\"a}nggi}\ and\ \citenamefont
  {Thomas}(1975)}]{Haenggi:1975}%
  \BibitemOpen
  \bibfield  {author} {\bibinfo {author} {\bibfnamefont {P.}~\bibnamefont
  {H{\"a}nggi}}\ and\ \bibinfo {author} {\bibfnamefont {H.}~\bibnamefont
  {Thomas}},\ }\bibfield  {title} {\bibinfo {title} {Linear response and
  fluctuation theorems for nonstationary stochastic processes},\ }\href
  {https://doi.org/10.1007/BF01362253} {\bibfield  {journal} {\bibinfo
  {journal} {Z. Phys. B}\ }\textbf {\bibinfo {volume} {22}},\ \bibinfo {pages}
  {295} (\bibinfo {year} {1975})}\BibitemShut {NoStop}%
\bibitem [{\citenamefont {Prost}\ \emph {et~al.}(2009)\citenamefont {Prost},
  \citenamefont {Joanny},\ and\ \citenamefont {Parrondo}}]{Prost:2009}%
  \BibitemOpen
  \bibfield  {author} {\bibinfo {author} {\bibfnamefont {J.}~\bibnamefont
  {Prost}}, \bibinfo {author} {\bibfnamefont {J.-F.}\ \bibnamefont {Joanny}},\
  and\ \bibinfo {author} {\bibfnamefont {J.~M.~R.}\ \bibnamefont {Parrondo}},\
  }\bibfield  {title} {\bibinfo {title} {Generalized fluctuation-dissipation
  theorem for steady-state systems},\ }\href
  {https://doi.org/10.1103/PhysRevLett.103.090601} {\bibfield  {journal}
  {\bibinfo  {journal} {Phys. Rev. Lett.}\ }\textbf {\bibinfo {volume} {103}},\
  \bibinfo {pages} {090601} (\bibinfo {year} {2009})}\BibitemShut {NoStop}%
\bibitem [{\citenamefont {Chiocchetta}\ \emph {et~al.}(2017)\citenamefont
  {Chiocchetta}, \citenamefont {Gambassi},\ and\ \citenamefont
  {Carusotto}}]{Chiocchetta:2017b}%
  \BibitemOpen
  \bibfield  {author} {\bibinfo {author} {\bibfnamefont {A.}~\bibnamefont
  {Chiocchetta}}, \bibinfo {author} {\bibfnamefont {A.}~\bibnamefont
  {Gambassi}},\ and\ \bibinfo {author} {\bibfnamefont {I.}~\bibnamefont
  {Carusotto}},\ }\bibinfo {title} {Laser operation and {B}ose-{E}instein
  condensation: {A}nalogies and differences},\ in\ \href
  {https://doi.org/10.1017/9781316084366.022} {\emph {\bibinfo {booktitle}
  {Universal {T}hemes of {B}ose-{E}instein {C}ondensation}}},\ \bibinfo
  {editor} {edited by\ \bibinfo {editor} {\bibfnamefont {N.~P.}\ \bibnamefont
  {Proukakis}}, \bibinfo {editor} {\bibfnamefont {D.~W.}\ \bibnamefont
  {Snoke}},\ and\ \bibinfo {editor} {\bibfnamefont {P.~B.}\ \bibnamefont
  {Littlewood}}}\ (\bibinfo  {publisher} {{C}ambridge {U}niversity {P}ress},\
  \bibinfo {year} {2017})\ pp.\ \bibinfo {pages} {409–423,
  \href{https://arxiv.org/abs/1503.02816}{ arXiv:1503.02816}}\BibitemShut
  {NoStop}%
\bibitem [{\citenamefont {Askitopoulos}\ \emph {et~al.}(2013)\citenamefont
  {Askitopoulos}, \citenamefont {Ohadi}, \citenamefont {Kavokin}, \citenamefont
  {Hatzopoulos}, \citenamefont {Savvidis},\ and\ \citenamefont
  {Lagoudakis}}]{Askitopoulos:2013}%
  \BibitemOpen
  \bibfield  {author} {\bibinfo {author} {\bibfnamefont {A.}~\bibnamefont
  {Askitopoulos}}, \bibinfo {author} {\bibfnamefont {H.}~\bibnamefont {Ohadi}},
  \bibinfo {author} {\bibfnamefont {A.~V.}\ \bibnamefont {Kavokin}}, \bibinfo
  {author} {\bibfnamefont {Z.}~\bibnamefont {Hatzopoulos}}, \bibinfo {author}
  {\bibfnamefont {P.~G.}\ \bibnamefont {Savvidis}},\ and\ \bibinfo {author}
  {\bibfnamefont {P.~G.}\ \bibnamefont {Lagoudakis}},\ }\bibfield  {title}
  {\bibinfo {title} {Polariton condensation in an optically induced
  two-dimensional potential},\ }\href
  {https://doi.org/10.1103/PhysRevB.88.041308} {\bibfield  {journal} {\bibinfo
  {journal} {Phys. Rev. B}\ }\textbf {\bibinfo {volume} {88}},\ \bibinfo
  {pages} {041308} (\bibinfo {year} {2013})}\BibitemShut {NoStop}%
\bibitem [{\citenamefont {Schmitt}\ \emph {et~al.}(2014)\citenamefont
  {Schmitt}, \citenamefont {Damm}, \citenamefont {Dung}, \citenamefont
  {Vewinger}, \citenamefont {Klaers},\ and\ \citenamefont
  {Weitz}}]{Schmitt:2014}%
  \BibitemOpen
  \bibfield  {author} {\bibinfo {author} {\bibfnamefont {J.}~\bibnamefont
  {Schmitt}}, \bibinfo {author} {\bibfnamefont {T.}~\bibnamefont {Damm}},
  \bibinfo {author} {\bibfnamefont {D.}~\bibnamefont {Dung}}, \bibinfo {author}
  {\bibfnamefont {F.}~\bibnamefont {Vewinger}}, \bibinfo {author}
  {\bibfnamefont {J.}~\bibnamefont {Klaers}},\ and\ \bibinfo {author}
  {\bibfnamefont {M.}~\bibnamefont {Weitz}},\ }\bibfield  {title} {\bibinfo
  {title} {Observation of grand-canonical number statistics in a photon
  {B}ose-{E}instein condensate},\ }\href
  {https://doi.org/10.1103/PhysRevLett.112.030401} {\bibfield  {journal}
  {\bibinfo  {journal} {Phys. Rev. Lett.}\ }\textbf {\bibinfo {volume} {112}},\
  \bibinfo {pages} {030401} (\bibinfo {year} {2014})}\BibitemShut {NoStop}%
\bibitem [{\citenamefont {Schmitt}\ \emph {et~al.}(2016)\citenamefont
  {Schmitt}, \citenamefont {Damm}, \citenamefont {Dung}, \citenamefont {Wahl},
  \citenamefont {Vewinger}, \citenamefont {Klaers},\ and\ \citenamefont
  {Weitz}}]{Schmitt:2016}%
  \BibitemOpen
  \bibfield  {author} {\bibinfo {author} {\bibfnamefont {J.}~\bibnamefont
  {Schmitt}}, \bibinfo {author} {\bibfnamefont {T.}~\bibnamefont {Damm}},
  \bibinfo {author} {\bibfnamefont {D.}~\bibnamefont {Dung}}, \bibinfo {author}
  {\bibfnamefont {C.}~\bibnamefont {Wahl}}, \bibinfo {author} {\bibfnamefont
  {F.}~\bibnamefont {Vewinger}}, \bibinfo {author} {\bibfnamefont
  {J.}~\bibnamefont {Klaers}},\ and\ \bibinfo {author} {\bibfnamefont
  {M.}~\bibnamefont {Weitz}},\ }\bibfield  {title} {\bibinfo {title}
  {Spontaneous symmetry breaking and phase coherence of a photon
  {B}ose-{E}instein condensate coupled to a reservoir},\ }\href
  {https://doi.org/10.1103/PhysRevLett.116.033604} {\bibfield  {journal}
  {\bibinfo  {journal} {Phys. Rev. Lett.}\ }\textbf {\bibinfo {volume} {116}},\
  \bibinfo {pages} {033604} (\bibinfo {year} {2016})}\BibitemShut {NoStop}%
\bibitem [{\citenamefont {Baboux}\ \emph {et~al.}(2018)\citenamefont {Baboux},
  \citenamefont {Bernardis}, \citenamefont {Goblot}, \citenamefont {Gladilin},
  \citenamefont {Gomez}, \citenamefont {Galopin}, \citenamefont {Gratiet},
  \citenamefont {Lema\^{i}tre}, \citenamefont {Sagnes}, \citenamefont
  {Carusotto}, \citenamefont {Wouters}, \citenamefont {Amo},\ and\
  \citenamefont {Bloch}}]{Baboux:2018}%
  \BibitemOpen
  \bibfield  {author} {\bibinfo {author} {\bibfnamefont {F.}~\bibnamefont
  {Baboux}}, \bibinfo {author} {\bibfnamefont {D.~D.}\ \bibnamefont
  {Bernardis}}, \bibinfo {author} {\bibfnamefont {V.}~\bibnamefont {Goblot}},
  \bibinfo {author} {\bibfnamefont {V.~N.}\ \bibnamefont {Gladilin}}, \bibinfo
  {author} {\bibfnamefont {C.}~\bibnamefont {Gomez}}, \bibinfo {author}
  {\bibfnamefont {E.}~\bibnamefont {Galopin}}, \bibinfo {author} {\bibfnamefont
  {L.~L.}\ \bibnamefont {Gratiet}}, \bibinfo {author} {\bibfnamefont
  {A.}~\bibnamefont {Lema\^{i}tre}}, \bibinfo {author} {\bibfnamefont
  {I.}~\bibnamefont {Sagnes}}, \bibinfo {author} {\bibfnamefont
  {I.}~\bibnamefont {Carusotto}}, \bibinfo {author} {\bibfnamefont
  {M.}~\bibnamefont {Wouters}}, \bibinfo {author} {\bibfnamefont
  {A.}~\bibnamefont {Amo}},\ and\ \bibinfo {author} {\bibfnamefont
  {J.}~\bibnamefont {Bloch}},\ }\bibfield  {title} {\bibinfo {title} {Unstable
  and stable regimes of polariton condensation},\ }\href
  {https://doi.org/10.1364/OPTICA.5.001163} {\bibfield  {journal} {\bibinfo
  {journal} {Optica}\ }\textbf {\bibinfo {volume} {5}},\ \bibinfo {pages}
  {1163} (\bibinfo {year} {2018})}\BibitemShut {NoStop}%
\bibitem [{\citenamefont {Pieczarka}\ \emph {et~al.}(2020)\citenamefont
  {Pieczarka}, \citenamefont {Estrecho}, \citenamefont {Boozarjmehr},
  \citenamefont {Bleu}, \citenamefont {Steger}, \citenamefont {West},
  \citenamefont {Pfeiffer}, \citenamefont {Snoke}, \citenamefont {Levinsen},
  \citenamefont {Parish}, \citenamefont {Truscott},\ and\ \citenamefont
  {Ostrovskaya}}]{Pieczarka:2020}%
  \BibitemOpen
  \bibfield  {author} {\bibinfo {author} {\bibfnamefont {M.}~\bibnamefont
  {Pieczarka}}, \bibinfo {author} {\bibfnamefont {E.}~\bibnamefont {Estrecho}},
  \bibinfo {author} {\bibfnamefont {M.}~\bibnamefont {Boozarjmehr}}, \bibinfo
  {author} {\bibfnamefont {O.}~\bibnamefont {Bleu}}, \bibinfo {author}
  {\bibfnamefont {M.}~\bibnamefont {Steger}}, \bibinfo {author} {\bibfnamefont
  {K.}~\bibnamefont {West}}, \bibinfo {author} {\bibfnamefont {L.~N.}\
  \bibnamefont {Pfeiffer}}, \bibinfo {author} {\bibfnamefont {D.~W.}\
  \bibnamefont {Snoke}}, \bibinfo {author} {\bibfnamefont {J.}~\bibnamefont
  {Levinsen}}, \bibinfo {author} {\bibfnamefont {M.~M.}\ \bibnamefont
  {Parish}}, \bibinfo {author} {\bibfnamefont {A.~G.}\ \bibnamefont
  {Truscott}},\ and\ \bibinfo {author} {\bibfnamefont {E.~A.}\ \bibnamefont
  {Ostrovskaya}},\ }\bibfield  {title} {\bibinfo {title} {Observation of
  quantum depletion in a non-equilibrium exciton--polariton condensate},\
  }\href {https://doi.org/10.1038/s41467-019-14243-6} {\bibfield  {journal}
  {\bibinfo  {journal} {Nat. Comm.}\ }\textbf {\bibinfo {volume} {11}},\
  \bibinfo {pages} {429} (\bibinfo {year} {2020})}\BibitemShut {NoStop}%
\bibitem [{\citenamefont {Wang}\ \emph {et~al.}(2021)\citenamefont {Wang},
  \citenamefont {Sigurdsson}, \citenamefont {T\"opfer},\ and\ \citenamefont
  {Lagoudakis}}]{Wang:2021}%
  \BibitemOpen
  \bibfield  {author} {\bibinfo {author} {\bibfnamefont {Y.}~\bibnamefont
  {Wang}}, \bibinfo {author} {\bibfnamefont {H.}~\bibnamefont {Sigurdsson}},
  \bibinfo {author} {\bibfnamefont {J.~D.}\ \bibnamefont {T\"opfer}},\ and\
  \bibinfo {author} {\bibfnamefont {P.~G.}\ \bibnamefont {Lagoudakis}},\
  }\bibfield  {title} {\bibinfo {title} {Reservoir optics with
  exciton-polariton condensates},\ }\href
  {https://doi.org/10.1103/PhysRevB.104.235306} {\bibfield  {journal} {\bibinfo
   {journal} {Phys. Rev. B}\ }\textbf {\bibinfo {volume} {104}},\ \bibinfo
  {pages} {235306} (\bibinfo {year} {2021})}\BibitemShut {NoStop}%
\bibitem [{\citenamefont {Klaers}\ \emph {et~al.}(2010)\citenamefont {Klaers},
  \citenamefont {Schmitt}, \citenamefont {Vewinger},\ and\ \citenamefont
  {Weitz}}]{Klaers:2010}%
  \BibitemOpen
  \bibfield  {author} {\bibinfo {author} {\bibfnamefont {J.}~\bibnamefont
  {Klaers}}, \bibinfo {author} {\bibfnamefont {J.}~\bibnamefont {Schmitt}},
  \bibinfo {author} {\bibfnamefont {F.}~\bibnamefont {Vewinger}},\ and\
  \bibinfo {author} {\bibfnamefont {M.}~\bibnamefont {Weitz}},\ }\bibfield
  {title} {\bibinfo {title} {{B}ose--{E}instein condensation of photons in an
  optical microcavity},\ }\href {https://doi.org/10.1038/nature09567}
  {\bibfield  {journal} {\bibinfo  {journal} {Nature}\ }\textbf {\bibinfo
  {volume} {468}},\ \bibinfo {pages} {545} (\bibinfo {year}
  {2010})}\BibitemShut {NoStop}%
\bibitem [{\citenamefont {Marelic}\ and\ \citenamefont
  {Nyman}(2015)}]{Marelic:2015}%
  \BibitemOpen
  \bibfield  {author} {\bibinfo {author} {\bibfnamefont {J.}~\bibnamefont
  {Marelic}}\ and\ \bibinfo {author} {\bibfnamefont {R.~A.}\ \bibnamefont
  {Nyman}},\ }\bibfield  {title} {\bibinfo {title} {Experimental evidence for
  inhomogeneous pumping and energy-dependent effects in photon
  {B}ose-{E}instein condensation},\ }\href
  {https://doi.org/10.1103/PhysRevA.91.033813} {\bibfield  {journal} {\bibinfo
  {journal} {Phys. Rev. A}\ }\textbf {\bibinfo {volume} {91}},\ \bibinfo
  {pages} {033813} (\bibinfo {year} {2015})}\BibitemShut {NoStop}%
\bibitem [{\citenamefont {Greveling}\ \emph {et~al.}(2018)\citenamefont
  {Greveling}, \citenamefont {Perrier},\ and\ \citenamefont {van
  Oosten}}]{Greveling:2018}%
  \BibitemOpen
  \bibfield  {author} {\bibinfo {author} {\bibfnamefont {S.}~\bibnamefont
  {Greveling}}, \bibinfo {author} {\bibfnamefont {K.~L.}\ \bibnamefont
  {Perrier}},\ and\ \bibinfo {author} {\bibfnamefont {D.}~\bibnamefont {van
  Oosten}},\ }\bibfield  {title} {\bibinfo {title} {Density distribution of a
  {B}ose-{E}instein condensate of photons in a dye-filled microcavity},\ }\href
  {https://doi.org/10.1103/PhysRevA.98.013810} {\bibfield  {journal} {\bibinfo
  {journal} {Phys. Rev. A}\ }\textbf {\bibinfo {volume} {98}},\ \bibinfo
  {pages} {013810} (\bibinfo {year} {2018})}\BibitemShut {NoStop}%
\bibitem [{\citenamefont {Ziff}\ \emph {et~al.}(1977)\citenamefont {Ziff},
  \citenamefont {Uhlenbeck},\ and\ \citenamefont {Kac}}]{Ziff:1977}%
  \BibitemOpen
  \bibfield  {author} {\bibinfo {author} {\bibfnamefont {R.~M.}\ \bibnamefont
  {Ziff}}, \bibinfo {author} {\bibfnamefont {G.~E.}\ \bibnamefont
  {Uhlenbeck}},\ and\ \bibinfo {author} {\bibfnamefont {M.}~\bibnamefont
  {Kac}},\ }\bibfield  {title} {\bibinfo {title} {The ideal {B}ose-{E}instein
  gas, revisited},\ }\href
  {https://doi.org/https://doi.org/10.1016/0370-1573(77)90052-7} {\bibfield
  {journal} {\bibinfo  {journal} {Phys. Rep.}\ }\textbf {\bibinfo {volume}
  {32}},\ \bibinfo {pages} {169} (\bibinfo {year} {1977})}\BibitemShut
  {NoStop}%
\bibitem [{\citenamefont {Klaers}\ \emph {et~al.}(2012)\citenamefont {Klaers},
  \citenamefont {Schmitt}, \citenamefont {Damm}, \citenamefont {Vewinger},\
  and\ \citenamefont {Weitz}}]{Klaers:2012}%
  \BibitemOpen
  \bibfield  {author} {\bibinfo {author} {\bibfnamefont {J.}~\bibnamefont
  {Klaers}}, \bibinfo {author} {\bibfnamefont {J.}~\bibnamefont {Schmitt}},
  \bibinfo {author} {\bibfnamefont {T.}~\bibnamefont {Damm}}, \bibinfo {author}
  {\bibfnamefont {F.}~\bibnamefont {Vewinger}},\ and\ \bibinfo {author}
  {\bibfnamefont {M.}~\bibnamefont {Weitz}},\ }\bibfield  {title} {\bibinfo
  {title} {Statistical physics of {B}ose-{E}instein-condensed light in a dye
  microcavity},\ }\href {https://doi.org/10.1103/PhysRevLett.108.160403}
  {\bibfield  {journal} {\bibinfo  {journal} {Phys. Rev. Lett.}\ }\textbf
  {\bibinfo {volume} {108}},\ \bibinfo {pages} {160403} (\bibinfo {year}
  {2012})}\BibitemShut {NoStop}%
\bibitem [{\citenamefont {Sob'yanin}(2012)}]{Sobyanin:2012}%
  \BibitemOpen
  \bibfield  {author} {\bibinfo {author} {\bibfnamefont {D.~N.}\ \bibnamefont
  {Sob'yanin}},\ }\bibfield  {title} {\bibinfo {title} {Hierarchical maximum
  entropy principle for generalized superstatistical systems and
  {B}ose-{E}instein condensation of light},\ }\href
  {https://doi.org/10.1103/PhysRevE.85.061120} {\bibfield  {journal} {\bibinfo
  {journal} {Phys. Rev. E}\ }\textbf {\bibinfo {volume} {85}},\ \bibinfo
  {pages} {061120} (\bibinfo {year} {2012})}\BibitemShut {NoStop}%
\bibitem [{\citenamefont {Gladilin}\ and\ \citenamefont
  {Wouters}(2020)}]{Gladilin:2020a}%
  \BibitemOpen
  \bibfield  {author} {\bibinfo {author} {\bibfnamefont {V.~N.}\ \bibnamefont
  {Gladilin}}\ and\ \bibinfo {author} {\bibfnamefont {M.}~\bibnamefont
  {Wouters}},\ }\bibfield  {title} {\bibinfo {title} {Classical field model for
  arrays of photon condensates},\ }\href
  {https://doi.org/10.1103/PhysRevA.101.043814} {\bibfield  {journal} {\bibinfo
   {journal} {Phys. Rev. A}\ }\textbf {\bibinfo {volume} {101}},\ \bibinfo
  {pages} {043814} (\bibinfo {year} {2020})}\BibitemShut {NoStop}%
\bibitem [{\citenamefont {Deng}\ \emph {et~al.}(2006)\citenamefont {Deng},
  \citenamefont {Press}, \citenamefont {G\"otzinger}, \citenamefont {Solomon},
  \citenamefont {Hey}, \citenamefont {Ploog},\ and\ \citenamefont
  {Yamamoto}}]{Deng:2006}%
  \BibitemOpen
  \bibfield  {author} {\bibinfo {author} {\bibfnamefont {H.}~\bibnamefont
  {Deng}}, \bibinfo {author} {\bibfnamefont {D.}~\bibnamefont {Press}},
  \bibinfo {author} {\bibfnamefont {S.}~\bibnamefont {G\"otzinger}}, \bibinfo
  {author} {\bibfnamefont {G.~S.}\ \bibnamefont {Solomon}}, \bibinfo {author}
  {\bibfnamefont {R.}~\bibnamefont {Hey}}, \bibinfo {author} {\bibfnamefont
  {K.~H.}\ \bibnamefont {Ploog}},\ and\ \bibinfo {author} {\bibfnamefont
  {Y.}~\bibnamefont {Yamamoto}},\ }\bibfield  {title} {\bibinfo {title}
  {Quantum degenerate {E}xciton-{P}olaritons in thermal equilibrium},\ }\href
  {https://doi.org/10.1103/PhysRevLett.97.146402} {\bibfield  {journal}
  {\bibinfo  {journal} {Phys. Rev. Lett.}\ }\textbf {\bibinfo {volume} {97}},\
  \bibinfo {pages} {146402} (\bibinfo {year} {2006})}\BibitemShut {NoStop}%
\bibitem [{\citenamefont {Schmitt}\ \emph {et~al.}(2015)\citenamefont
  {Schmitt}, \citenamefont {Damm}, \citenamefont {Dung}, \citenamefont
  {Vewinger}, \citenamefont {Klaers},\ and\ \citenamefont
  {Weitz}}]{Schmitt:2015}%
  \BibitemOpen
  \bibfield  {author} {\bibinfo {author} {\bibfnamefont {J.}~\bibnamefont
  {Schmitt}}, \bibinfo {author} {\bibfnamefont {T.}~\bibnamefont {Damm}},
  \bibinfo {author} {\bibfnamefont {D.}~\bibnamefont {Dung}}, \bibinfo {author}
  {\bibfnamefont {F.}~\bibnamefont {Vewinger}}, \bibinfo {author}
  {\bibfnamefont {J.}~\bibnamefont {Klaers}},\ and\ \bibinfo {author}
  {\bibfnamefont {M.}~\bibnamefont {Weitz}},\ }\bibfield  {title} {\bibinfo
  {title} {Thermalization kinetics of light: From laser dynamics to equilibrium
  condensation of photons},\ }\href
  {https://doi.org/10.1103/PhysRevA.92.011602} {\bibfield  {journal} {\bibinfo
  {journal} {Phys. Rev. A}\ }\textbf {\bibinfo {volume} {92}},\ \bibinfo
  {pages} {011602} (\bibinfo {year} {2015})}\BibitemShut {NoStop}%
\bibitem [{\citenamefont {Sun}\ \emph {et~al.}(2017)\citenamefont {Sun},
  \citenamefont {Wen}, \citenamefont {Yoon}, \citenamefont {Liu}, \citenamefont
  {Steger}, \citenamefont {Pfeiffer}, \citenamefont {West}, \citenamefont
  {Snoke},\ and\ \citenamefont {Nelson}}]{Sun:2017}%
  \BibitemOpen
  \bibfield  {author} {\bibinfo {author} {\bibfnamefont {Y.}~\bibnamefont
  {Sun}}, \bibinfo {author} {\bibfnamefont {P.}~\bibnamefont {Wen}}, \bibinfo
  {author} {\bibfnamefont {Y.}~\bibnamefont {Yoon}}, \bibinfo {author}
  {\bibfnamefont {G.}~\bibnamefont {Liu}}, \bibinfo {author} {\bibfnamefont
  {M.}~\bibnamefont {Steger}}, \bibinfo {author} {\bibfnamefont {L.~N.}\
  \bibnamefont {Pfeiffer}}, \bibinfo {author} {\bibfnamefont {K.}~\bibnamefont
  {West}}, \bibinfo {author} {\bibfnamefont {D.~W.}\ \bibnamefont {Snoke}},\
  and\ \bibinfo {author} {\bibfnamefont {K.~A.}\ \bibnamefont {Nelson}},\
  }\bibfield  {title} {\bibinfo {title} {Bose-einstein condensation of
  long-lifetime polaritons in thermal equilibrium},\ }\href
  {https://doi.org/10.1103/PhysRevLett.118.016602} {\bibfield  {journal}
  {\bibinfo  {journal} {Phys. Rev. Lett.}\ }\textbf {\bibinfo {volume} {118}},\
  \bibinfo {pages} {016602} (\bibinfo {year} {2017})}\BibitemShut {NoStop}%
\bibitem [{\citenamefont {Cugliandolo}(2011)}]{Cugliandolo:2011}%
  \BibitemOpen
  \bibfield  {author} {\bibinfo {author} {\bibfnamefont {L.~F.}\ \bibnamefont
  {Cugliandolo}},\ }\bibfield  {title} {\bibinfo {title} {The effective
  temperature},\ }\href {https://doi.org/10.1088/1751-8113/44/48/483001}
  {\bibfield  {journal} {\bibinfo  {journal} {J. Phys. A Math. Theor.}\
  }\textbf {\bibinfo {volume} {44}},\ \bibinfo {pages} {483001} (\bibinfo
  {year} {2011})}\BibitemShut {NoStop}%
\bibitem [{\citenamefont {Foini}\ \emph {et~al.}(2011)\citenamefont {Foini},
  \citenamefont {Cugliandolo},\ and\ \citenamefont {Gambassi}}]{Foini:2011}%
  \BibitemOpen
  \bibfield  {author} {\bibinfo {author} {\bibfnamefont {L.}~\bibnamefont
  {Foini}}, \bibinfo {author} {\bibfnamefont {L.~F.}\ \bibnamefont
  {Cugliandolo}},\ and\ \bibinfo {author} {\bibfnamefont {A.}~\bibnamefont
  {Gambassi}},\ }\bibfield  {title} {\bibinfo {title} {Fluctuation-dissipation
  relations and critical quenches in the transverse field ising chain},\ }\href
  {https://doi.org/10.1103/PhysRevB.84.212404} {\bibfield  {journal} {\bibinfo
  {journal} {Phys. Rev. B}\ }\textbf {\bibinfo {volume} {84}},\ \bibinfo
  {pages} {212404} (\bibinfo {year} {2011})}\BibitemShut {NoStop}%
\bibitem [{\citenamefont {Foini}\ \emph {et~al.}(2012)\citenamefont {Foini},
  \citenamefont {Cugliandolo},\ and\ \citenamefont {Gambassi}}]{Foini:2012}%
  \BibitemOpen
  \bibfield  {author} {\bibinfo {author} {\bibfnamefont {L.}~\bibnamefont
  {Foini}}, \bibinfo {author} {\bibfnamefont {L.~F.}\ \bibnamefont
  {Cugliandolo}},\ and\ \bibinfo {author} {\bibfnamefont {A.}~\bibnamefont
  {Gambassi}},\ }\bibfield  {title} {\bibinfo {title} {Dynamic correlations,
  fluctuation-dissipation relations, and effective temperatures after a quantum
  quench of the transverse field {I}sing chain},\ }\href
  {https://doi.org/10.1088/1742-5468/2012/09/p09011} {\bibfield  {journal}
  {\bibinfo  {journal} {J. Stat. Mech.}\ }\textbf {\bibinfo {volume} {2012}},\
  \bibinfo {pages} {P09011} (\bibinfo {year} {2012})}\BibitemShut {NoStop}%
\bibitem [{\citenamefont {Schmitt}(2018)}]{Schmitt:2018}%
  \BibitemOpen
  \bibfield  {author} {\bibinfo {author} {\bibfnamefont {J.}~\bibnamefont
  {Schmitt}},\ }\bibfield  {title} {\bibinfo {title} {Dynamics and correlations
  of a {B}ose{\textendash}{E}instein condensate of photons},\ }\href
  {https://doi.org/10.1088/1361-6455/aad409} {\bibfield  {journal} {\bibinfo
  {journal} {J. Phys. B: At. Mol. Opt. Phys.}\ }\textbf {\bibinfo {volume}
  {51}},\ \bibinfo {pages} {173001} (\bibinfo {year} {2018})}\BibitemShut
  {NoStop}%
\bibitem [{\citenamefont {\"Ozt\"urk}\ \emph {et~al.}(2021)\citenamefont
  {\"Ozt\"urk}, \citenamefont {Lappe}, \citenamefont {Hellmann}, \citenamefont
  {Schmitt}, \citenamefont {Klaers}, \citenamefont {Vewinger},\ and\
  \citenamefont {Weitz}}]{Oeztuerk:2021}%
  \BibitemOpen
  \bibfield  {author} {\bibinfo {author} {\bibfnamefont {F.~E.}\ \bibnamefont
  {\"Ozt\"urk}}, \bibinfo {author} {\bibfnamefont {T.}~\bibnamefont {Lappe}},
  \bibinfo {author} {\bibfnamefont {G.}~\bibnamefont {Hellmann}}, \bibinfo
  {author} {\bibfnamefont {J.}~\bibnamefont {Schmitt}}, \bibinfo {author}
  {\bibfnamefont {J.}~\bibnamefont {Klaers}}, \bibinfo {author} {\bibfnamefont
  {F.}~\bibnamefont {Vewinger}},\ and\ \bibinfo {author} {\bibfnamefont
  {M.}~\bibnamefont {Weitz}},\ }\bibfield  {title} {\bibinfo {title}
  {Observation of a non-{H}ermitian phase transition in an optical quantum
  gas},\ }\href {https://doi.org/10.1126/science.abe9869} {\bibfield  {journal}
  {\bibinfo  {journal} {Science}\ }\textbf {\bibinfo {volume} {372}},\ \bibinfo
  {pages} {88} (\bibinfo {year} {2021})}\BibitemShut {NoStop}%
\bibitem [{\citenamefont {Busley}\ \emph {et~al.}(2022)\citenamefont {Busley},
  \citenamefont {Miranda}, \citenamefont {Redmann}, \citenamefont {Kurtscheid},
  \citenamefont {Umesh}, \citenamefont {Vewinger}, \citenamefont {Weitz},\ and\
  \citenamefont {Schmitt}}]{Busley:2022}%
  \BibitemOpen
  \bibfield  {author} {\bibinfo {author} {\bibfnamefont {E.}~\bibnamefont
  {Busley}}, \bibinfo {author} {\bibfnamefont {L.~E.}\ \bibnamefont {Miranda}},
  \bibinfo {author} {\bibfnamefont {A.}~\bibnamefont {Redmann}}, \bibinfo
  {author} {\bibfnamefont {C.}~\bibnamefont {Kurtscheid}}, \bibinfo {author}
  {\bibfnamefont {K.~K.}\ \bibnamefont {Umesh}}, \bibinfo {author}
  {\bibfnamefont {F.}~\bibnamefont {Vewinger}}, \bibinfo {author}
  {\bibfnamefont {M.}~\bibnamefont {Weitz}},\ and\ \bibinfo {author}
  {\bibfnamefont {J.}~\bibnamefont {Schmitt}},\ }\bibfield  {title} {\bibinfo
  {title} {Compressibility and the equation of state of an optical quantum gas
  in a box},\ }\href {https://doi.org/10.1126/science.abm2543} {\bibfield
  {journal} {\bibinfo  {journal} {Science}\ }\textbf {\bibinfo {volume}
  {375}},\ \bibinfo {pages} {1403} (\bibinfo {year} {2022})}\BibitemShut
  {NoStop}%
\bibitem [{\citenamefont {Huang}(1987)}]{Huang:1987}%
  \BibitemOpen
  \bibfield  {author} {\bibinfo {author} {\bibfnamefont {K.}~\bibnamefont
  {Huang}},\ }\href@noop {} {\emph {\bibinfo {title} {Statistical Mechanics}}}\
  (\bibinfo  {publisher} {Wiley},\ \bibinfo {address} {New York},\ \bibinfo
  {year} {1987})\BibitemShut {NoStop}%
\end{thebibliography}%

\end{document}